# Substituent-Controlled Reversible Switching of Charge-Injection-Barrier Heights at Metal/Organic-Semiconductor Contacts Modified with Disordered Molecular Monolayers

*Ryo Nouchi\* and Takaaki Tanimoto*

Nanoscience and Nanotechnology Research Center, Osaka Prefecture University, Sakai 599-8570, Japan

ABSTRACT. Electrically stimulated switching of a charge injection barrier at the interface between an organic semiconductor and an electrode modified with a disordered monolayer (DM) is studied by using various benzenethiol derivatives as DM molecules. The switching behavior is induced by a structural change in the DM molecules, and is manifested as a reversible inversion of the polarity of DM-modified Au electrode/rubrene/DM-modified Au electrode diodes. The switching direction is found to be dominantly determined by the push-back effect of the thiol bonding group, while the terminal group modulates the switching strength. A device with 1,2-benzenedithiol DMs exhibited the highest switching ratios of 20, $10^2$, and $10^3$ for the switching voltages of 3, 5, and 7 V, respectively. A variation in the tilt angle of benzenethiol DMs owing to the application of 7 V is estimated to be smaller than 23.6° by model calculations. This study offers an understanding for obtaining highly stable operations of organic electronic devices, especially with molecular modification layers.

Organic or molecular semiconductors are regarded as promising materials for realizing low-cost, large-area fabrication of electronic devices.[1,2] In most organic electronics and optoelectronics applications, intrinsic semiconductors are used as active layers. In this case, the type of charge carriers (electrons or holes) exploited in the electronic and optoelectronic devices is determined by a charge injection barrier at the electrode-semiconductor interface. If the barrier for electron injection is lower than that for hole, the device is classified as n-type, where electrons are the charge carriers. Therefore, it is critical to control the height of the charge injection barrier, $\Phi_B$. In a first approximation, the value of $\Phi_B$ is determined by the relative energy difference between the Fermi level, $E_F$, of the metallic electrode and the molecular orbital of the semiconducting molecule. Once the preferred molecule for the semiconducting layer is determined, the work function of the electrode, $\Phi_m$, must be tuned for controlling $\Phi_B$.

Among various methodologies for tuning of $\Phi_m$, modification of electrode surfaces with well-ordered self-assembled molecular monolayers (SAMs) is widely employed in the research of organic electronics. This is partly because organic semiconductor layers can be formed under rather mild conditions by formation techniques such as solution-based coating,[3,4] vacuum deposition with a low sublimation temperature,[5] or simple lamination of single crystals.[6-8] In contrast, a significant damage can be easily inflicted on a molecular monolayer by the deposition of inorganic materials on it.[9] In addition, the constituent molecules of SAMs can be flexibly designed by using organic synthetic techniques to obtain terminal group(s) possessing permanent electric dipoles. The $\Phi_m$ values are partly determined by the electric double layer on the surface.[10] Because a SAM of molecules with permanent dipoles forms an electric double layer on the electrode surface, the electrode's $\Phi_m$ is altered by modifying the surface by adding the SAM.

The direction (decrease or increase) and the magnitude of the change can be tuned by controlling the orientations and magnitudes of the dipoles.

The SAM-based technique has been considered as a static control of $\Phi_m$, in which the SAM-modified value of $\Phi_m$ is captured by a fixed value. Recently, efforts have been made to make the SAM-modified electrodes switchable. Switching by using an optical stimulus has been achieved for photochromic SAM molecules.[11] An organic field-effect transistor with source/drain electrodes modified with azobenzene-based SAMs was reported to exhibit switchability, although the reported switching ratio $R_{sw}$ of the electric current was low (~2). Switching by using an electrical stimulus has also been reported for electrodes modified with disordered monolayers (DMs).[12] A well-ordered structure of SAMs on a solid surface is formed by the adsorption and subsequent diffusion of the constituent molecules on the surface. The ordered structure is stabilized by the interconstituent van der Waals interactions that make the structure rigid. Instead, the DMs were formed by hindering the surface diffusion by using molecules with multiple bonding groups, which significantly reduced the structure rigidity. As a result, $R_{sw}$ as high as $10^5$ was observed with DM-modified electrode/organic semiconductor/DM-modified electrode diodes.[12]

While the former approach based on an optical stimulus is important in molecular optoelectronics, the latter technique based on an electrical stimulus should be important in molecular electronics because it follows the basic concept of electronics: electrical control of an electrical signal. The electrical switching with $R_{sw}$ as high as $10^5$ after application of 30 V was observed with DMs of 1,12-dimethyl-5,8-[4]helicenedithiol.[12] The constituent molecule of these DMs possessed a helical structure with a skeleton of four fused benzene rings. Owing to its rather complicated molecular structure, it was unclear which part of the molecule determined the switching strength. In addition, a control experiment with monothiol counterparts should have been performed for determining the effectiveness of the strategy employing DMs with multiple bonding groups. Furthermore, an external voltage of 30 V was used for inducing the switching; however, the minimal voltage required for inducing the switching is not yet known.

In this study, a series of benzenethiol derivatives were used as DMs of the electrode surface, and an electrical switching behavior of the two-terminal planar devices, where an organic semiconductor layer was bridged over two modified electrodes (Figure 1A), was characterized in terms of the direction and strength of switching. The benzenethiol derivatives have a rather simple structure, and a variety of derivatives are commercially available. In this study, benzenethiol (BT), 4-methylbenzenethiol (MBT), 4-nitrobenzenethiol (NBT), and 1,2-benzenedithiol (B2T) molecules were employed (Figure 1B) for investigating the effects of permanent electric dipoles of the terminal groups as well as the effects of the number of bonding groups. The results showed that the bonding group dominantly determines the switching direction, while the terminal group modifies the switching strength. In addition, the strategy for using molecules with multiple bonding groups was confirmed to be effective for inducing large-magnitude electrical switching.

## **RESULTS**

Figure 2A shows the current-voltage ($I$-$V$) characteristics of the as-fabricated two-terminal device with B2T-modified electrodes. The characteristics were measured in ambient air, under ambient light, and at room temperature. Because the as-fabricated device had a symmetric structure (a B2T-modified electrode/rubrene single crystal/B2T-modified electrode), the current levels of the initial *I-V* curve in the positively and negatively biased regions were naturally measured to be in the same order. Then, a voltage higher than that used for measuring the *I-V* curve (*i.e.*, up to ±1 V) was applied to the device for ~60 s to induce a structural switching of the B2T monolayer. Here, the switching voltage, $V_{sw}$, was set to +7 V. To determine whether the switching had been induced, the *I-V* characteristics were measured again, and the results are shown in Figure 2B. The *I-V* curve is clearly asymmetric, *i.e.*, it exhibits the rectification behavior of a diode, which was not observed in the initial curve. Next, $V_{sw}$ with the opposite polarity (−7 V) was applied to the same device for ~60 s. The *I-V* curve measured after applying negative $V_{sw}$ is shown in Figure 2C. This *I-V* curve is also asymmetric, but the observed rectification polarity is opposite to that shown in Figure 2B.

The polarity reversal was repeatedly observed after consecutive applications of positive and negative $V_{sw}$'s. The absolute current |*I*| at ±1 V of the *I-V* curves measured after each $V_{sw}$ application is plotted in Figure 2D. The $R_{sw}$ value of the B2T device with |$V_{sw}$| of 7 V reached ~$10^3$. These results indicate that $V_{sw}$ induced reversible switching of the work function of B2T-modified electrodes.

Two possible types of carriers can flow through the device; the carrier type is determined by the relative magnitudes of the charge-injection barrier height for electrons and holes. The electron and hole injection barrier heights have been reported to be 1.57 and 1.10 eV, respectively, for a rubrene thin film deposited on a Au film.[13] The barrier heights can be changed to 2.17 and 0.50 eV, respectively, by accounting for the band bending of 0.60 eV at the interface.[14] Thus, the height of the hole injection barrier is lower than that of the electron injection at the interface with the bare Au film. The work functions of Au films modified with BT derivatives should be different from that of the bare Au film. Modification by NBT should increase the work function, making the hole injection more efficient than electron injection. MBT (BT) was reported to lower the work function of a Au film (a Au(111) surface) by 0.36 eV (Ref. 15) (0.60 eV (Ref. 16)), but the estimated hole injection barrier of 0.86 eV (1.10 eV) is still lower than the estimated electron injection barrier of 1.81 eV (1.57 eV). The work function change of a Cu(110) surface by B2T adsorption was found to be 0.10 eV lower than that by BT adsorption.[17] If this difference holds in the case of Au surfaces, the estimated hole injection barrier of 1.00 eV is again lower than the estimated electron injection barrier of 1.67 eV. Therefore, in all of the combinations investigated in the present study, the electric current $I$ was considered to be based on the hole transport through the highest occupied molecular orbital (HOMO) of rubrene.

The schematic band diagrams corresponding to the $I$-$V$ curves in Figures 2A–C can be deduced from the $V_{sw}$-induced changes in the $I$-$V$ curves, as shown in Figures 2E–G, respectively. Positive and negative $V_{sw}$ enhance (impede) the current flow in the positively and negatively (negatively and positively) biased regions, respectively.

Figures 3A–C are the obtained switching cycles of the BT, MBT, and NBT devices for $V_{sw}$ = ±7 V. For all of the devices, the switching directions are the same as for the B2T device (Figure 2D). The switching cycles of the devices were examined for various $V_{sw}$ values of ±1, 3, 5, and 7 V. The $V_{sw}$ dependence of the average $R_{sw}$ is shown in Figure 3D along with the data for the devices with other BT derivatives. Compared with the BT device, the magnitude of the switching was higher for the B2T device, comparable for the MBT device, and lower for the NBT device. The control experiments with no DM revealed a very weak switching behavior, confirming that the above features originate from the presence of the DMs (see Supporting Information).

**DISCUSSION**

**Electric-field-induced structural change of DMs.** The reversible polarity switching of organic-semiconductor-based diodes discussed in this study is reminiscent of resistive switching phenomena that have been reported with various transition metal oxides (TMOs).[18] However, proposed models for the TMO-based switching can be likely excluded, and a structural change of the molecular monolayer formed on Au electrodes was proposed instead in the previous manuscript (see supplementary material of Ref. 12 for details). If the monolayer molecules possess permanent electrical dipole(s), an external electric field couples with the dipole charges to exert a Coulomb force on the monolayer. This field-dipole coupling has been shown to induce a structural change of a molecular monolayer on Au electrodes.[19]

The structural change of the DMs on the Au electrodes requires the lifting of the rubrene single crystal formed on the DMs. Such cargo-lifting phenomenon has been observed with azobenzene monolayers formed on Au films, where the monolayers electrically contacted by a Hg drop on it reversibly exhibited a structural change upon photo-irradiation even with the heavy Hg drop ($\sim 10^5$ N/m$^2$).[20] In our system, the upper layer pressure is at most 0.06 N/m$^2$, based on the thickness of a rubrene single crystal used in this study (up to ~ 5 μm) and its mass density (1.26 g/cm$^3$ (Ref. 21)). In our system, the pressure on the monolayer is much lower than that reported for the azobenzene case. Although the switching mechanisms are different in these two cases (light-induced or electric-field-induced switching), the upper layer pressure in our system is considered to be sufficiently low for the monolayer to exhibit a structural change.

The switching behavior indicates that the device with the DMs possesses a certain degree of non-volatility. In the previous paper,[12] the device with 1,12-dimethyl-5,8-[4]helicenedithiol monolayers retained the diode behavior at least for two days following the switching although the rectification ratio (defined as the ratio of the absolute electric currents at ±1 V) decreased from $10^5$ to $10^2$. For the systems studied in the present paper, the van der Waals interaction between the molecular skeleton and the Au surface might have caused similar non-volatility, although the retention time was not investigated in the present system because the main purpose of the present study was to investigate the effect of the substituents on the switching strength/direction. The retention time might be increased by altering the molecular structure to increase the molecule/electrode-surface interaction, which should be an important task in the future.

**Switching directions of BT, MBT, and B2T.** In the case of BT, a permanent electric dipole exists at the thiol bonding group. The moment points toward the benzene ring from the sulfur atom and its magnitude is 1.22 D at 45°;[22] thus, the component normal to the electrode surface becomes 0.86 D. In the case of B2T, the magnitude becomes 1.18 D owing to the structure rotated by 30° (Figure 1B). When an external electric field with the direction toward the electrode surface is applied to the BT-modified electrode, the BT molecules are tilted owing to the electrical coupling of the dipoles of the BT molecules with the downward electric field. A static change in $\Phi_m$ upon the formation of a molecular monolayer is determined by the so-called push-back (or pillow) effect and the dipole effect.[23] Compared with the $\Phi_m$ value immediately after the static change, *i.e.*, before the tilting (Figure 4A), $\Phi_m$ should be different owing to a change in the magnitudes of the push-back and dipole effects (Figures 4B and 4C).

A change in $\Phi_m$, $\Delta\Phi_m$, owing to the dipole effect is expressed as:[24]

$$\Delta\Phi_{\mathrm{m}}^{\mathrm{dipole}} = \frac{qN\mu_{\perp}}{\varepsilon_0} = \frac{qN\mu_0 \cos(\theta)}{\varepsilon_0 \varepsilon^{\mathrm{eff}}}, \tag{1}$$

where $q$ is the elementary charge, $N$ is the surface density of the modification layer molecules, $\mu_{\perp}$ is the normal component of the dipole moment of a single molecule, $\varepsilon_0$ is the vacuum permittivity, and $\theta$ is the tilt angle of the molecule relative to the surface normal. $\mu_0$ is the dipole moment of the isolated molecule, and it becomes positive (negative) when the dipole points towards the electrode surface (away from the surface). The quantity $\varepsilon^{\mathrm{eff}}$ represents the effective relative permittivity that accounts for the mutual depolarization of adjacent molecules, which is owing to the screening of dipole charges by the π-conjugated molecular cores.[25,26] Equation (1) indicates that the magnitude of $\Delta\Phi_{\mathrm{m}}^{\mathrm{dipole}}$ becomes largest for upright standing molecules and smallest for flat-lying molecules on the surface. In the case of BT, the dipole moment points away from the surface, yielding negative $\Delta\Phi_{\mathrm{m}}^{\mathrm{dipole}}$ (Figure 4A). Therefore, the tilting (standing) of the molecules decreases (increases) the magnitude of the negative $\Delta\Phi_{\mathrm{m}}^{\mathrm{dipole}}$, indicating that $\Phi_{\mathrm{m}}$ should increase (decrease) owing to the dipole effect upon tilting (standing) (Figures 4B and 4C).

The magnitude of the push-back effect is requisitely dependent on the average distance between the molecular skeletons of the monolayer and the metal surface. The electrons spilled out from the metal are pushed back into the metal owing to the Pauli repulsion from the electron clouds of the molecular skeletons. The push-back effect reduces $\Phi_{\mathrm{m}}$ by weakening the strength of the surface electric double layer at the metal surface (Figure 4A). The number of electrons that are pushed back into the metal increases (decreases) as the distance decreases (increases). Thus, the tilting (standing) of the molecules reduces (increases) $\Phi_{\mathrm{m}}$ (Figures 4B and 4C).

These considerations indicate that the dipole and push-back effects compete if the molecule has a permanent electric dipole in the same direction as the BT (Figures 4B and 4C). This competition determines the switching direction in the BT, MBT and B2T devices. The experimental results shown in Figure 3 indicate that the switching direction in the three devices was the same, where the band diagrams after the application of positive $V_{sw}$ became the same as that of the BT device (Figure 2F). The positive $V_{sw}$ generates an electric field pointing outward from (toward) the electrode surface at the right (left) electrode in the configuration displayed in Figure 3. Thus, the situation at the surface of the right and left electrodes is captured by Figures 4B and 4C, respectively. By comparing the band diagrams shown in these figures with those deduced from the experimental results (Figure 2F), the magnitude of the push-back effect is suggested to be larger than that of the dipole effect. The larger contribution of the push-back effect is consistent with our previous observation using 1,12-dimethyl-5,8-[4]helicenedithiol DMs.[12]

**Switching direction of NBT.** Next, we consider the case in which the dipole of the DM molecule is in the opposite direction to that of BT, *i.e.*, the NBT case. The dipole moment points toward the electrode surface; thus, $\Delta\Phi_m$ is positive owing to the dipole effect (Figure 4D). Therefore, the standing (tilted) molecules increase (reduce) the magnitude of positive $\Delta\Phi_m$, indicating that $\Phi_m$ should increase (decrease) owing to the dipole effect upon standing (tilting) (Figures 4E and 4F). The push-back effect reduces $\Phi_m$ by weakening the strength of the surface electric double layer at the metal surface (Figure 4D). The number of electrons that are pushed back into the metal increases (decreases) as the distance decreases (increases). Thus, the standing (tilted) molecules increase (reduce) $\Phi_m$ (Figures 4E and 4F). From the above considerations, it can be concluded that the dipole and push-back effects should induce $\Delta\Phi_m$ in the same direction upon the structural change of the NBT DM, *i.e.*, no competition between the dipole and push-back effects is expected.

More importantly, the switching direction of the NBT device should be opposite to the push-back-dominated switching of the BT, MBT, and B2T devices (compare Figures 4B and 4E or Figures 4C and 4F). However, the experimentally observed switching directions were the same for all DM molecules, as shown in Figures 2 and 3. Thus, the above consideration regarding the NBT device should disregard the mechanism that yields the same switching direction. To resolve this unexpected result for the NBT device, we further consider the specific dipoles at bonding and terminal groups instead of the single (overall) dipole.

First, the BT and B2T have no terminal groups. Next, the MBT has a methyl group, and the magnitude of its dipole is 0.37 D.[22] The dipole moment is upright towards the methyl group from the benzene ring; thus, its direction is the same as that of the thiol bonding group. Therefore, the switching directions of all dipoles are the same, and the switching is again expressed as in Figures 4A–C. Finally, in the case of NBT, a permanent electric dipole exists at the nitro terminal group. As shown in Figure 1B, the total dipole moment normal to the electrode surface becomes 3.15 D in the direction opposite to that of other molecules. This is because the moment at the nitro group is upright towards the benzene ring from the nitrogen atom, and its magnitude is 4.01 D.[22]

Only in the case of NBT the permanent dipoles at the bonding and terminal groups are opposite to each other, as shown in Figure 5A. Thus, the structural change in response to an external electric field should be different from that of the single dipole picture shown in Figures 4E and 4F. Figures 5B and 5C show the structural change in the NBT following the electric field application; in these figures, the C-N bond at the terminal group is assumed to possess a certain degree of flexibility, and thus the structural changes at the bonding and terminal groups are considered independently. Among the four sources that induce $\Delta\Phi_\mathrm{m}$ (the dipole and push-back effects of the bonding and terminal groups), only the push-back effect of the thiol bonding group can explain the experimentally observed switching direction. If the C-N bond is not flexible, opposite to what is depicted in Figures 5B and 5C, then the tilt angles of the bonding and terminal groups are expressed by a single value. This rigid-bond scenario becomes similar to the single-dipole scenarios in Figure 4. In the rigid-bond scenario, the experimental results can only be explained by the push-back effect arising from the bonding group. This conclusion is likely to be reasonable because the magnitude of the push-back effect is determined by how much the electronic clouds of the molecule and Au overlap (see "*Modeling the switching strength*" section for details) and the bonding group is closer to the electrode surface than the terminal group. Thus, the magnitude of the push-back effect of the bonding group can be naturally larger than the other contributions.

**Comparison of the switching ratios.** Figure 3D compares the $R_\mathrm{sw}$ values for all of the devices. It should be noted here that the $\Phi_\mathrm{m}$ values of the initial states were different among these systems. However, the initial difference in $\Phi_\mathrm{m}$ is considered not to affect $R_\mathrm{sw}$, by the following reasoning: The current switching is achieved by the change in $\Phi_\mathrm{B}$. The current transport through the system can be treated as the thermionic emission of charge carriers from the electrode into the organic semiconductor layer, as:[27]

$$I = AA^*T^2 \exp\left(-\frac{q\Phi_B}{kT}\right)\left[\exp\left(\frac{q(V-IR_s)}{nkT}\right)-1\right]$$

$$\approx \begin{cases} AA^*T^2 \exp\left(-\frac{q\Phi_B}{kT}\right)\exp\left(\frac{q(V-IR_s)}{nkT}\right) \text{ (for highly positive } V), \\ -AA^*T^2 \exp\left(-\frac{q\Phi_B}{kT}\right) \text{ (for highly negative } V) \end{cases} \quad (2)$$

where $A$ is the cross-section of the current flow path, $A^*$ is the effective Richardson constant, $T$ is the absolute temperature, $k$ is the Boltzmann constant, $R_s$ is the series resistance mainly arising from the semiconductor's bulk resistance, and $n$ is the ideality factor of the thermionic emission behavior. The expression for $I$ for highly negative $V$ is called "the reverse saturation current". In the case of the DM-modified electrode/organic semiconductor/DM-modified electrode diodes, energy barriers at both electrode/semiconductor interfaces should be considered. *I-V* characteristics of this double-Schottky-type device are known to be different from Eq. (2).[28] If the two barriers have different heights, the absolute magnitude of the current |*I*| in the *V* region with one polarity (positive or negative) is higher than that in the region with the other polarity. The higher and lower |*I*| were found to be expressed by the reverse saturation current of a single barrier diode with the lower and higher $\Phi_B$, respectively.[28] The application of $V_{sw}$ reversibly switches the $\Phi_B$ values as the interface with higher (lower) $\Phi_B$ turns into that with lower (higher) $\Phi_B$. Therefore, $R_{sw}$ is determined by the difference between the higher and lower heights, $\delta\Phi_B \equiv \Phi_B^{high} - \Phi_B^{low}$, as:

$$R_{sw} = \frac{AA^*T^2 \exp\left(-\frac{q\Phi_B^{low}}{kT}\right)}{AA^*T^2 \exp\left(-\frac{q\Phi_B^{high}}{kT}\right)} = \exp\left(\frac{q\delta\Phi_B}{kT}\right). \quad (3)$$

As a result, $R_{sw}$ does not depend on the initial $\Phi_m$, allowing us to perform a direct comparison of the $R_{sw}$ values compiled in Figure 3D. Thus, the switching strengths are in the following order B2T > BT > MBT > NBT for $|V_{sw}|$ of 7 V. Figure 3D also shows the $\delta\Phi_B$ values calculated by using Eq. (3) from the $R_{sw}$ data.

The strong switching of the B2T device is attributed to the multiple bonding nature. The B2T molecule has two thiol bonding groups. Thus, the force exerted by the external electric field is stronger than that for other monothiol molecules, which induces stronger structural change in the DM. In addition, dithiol molecules are known to form DM with higher disorder than that of DM formed by their monothiol counterparts.[29] A well-ordered SAM structure is formed by the adsorption and subsequent surface diffusion of the molecules. Two binding groups enable the molecules to strongly bind to the electrode surface, inhibiting the surface diffusion of the molecules. The resultant disordered monolayer of dithiol molecules has a non-rigid structure compared with its monothiol counterpart because the monolayer structural rigidity is determined by intermolecular van der Waals interactions. Possibly owing to these two facts, the B2T device exhibited the highest $R_{sw}$. In the present device sizing, the B2T device attained $R_{sw} = 10^3$ for $V_{sw} = 7$ V, and reached $10^2$ for $V_{sw} = 3$ V.

The weakest switching, that of the NBT device, is attributed to the permanent electric dipole at the nitro terminal group. As discussed for Figure 5, the bonding and terminal groups of the NBT possess dipoles in opposite directions. The $\Delta\Phi_m$ induced by the dipole effect of the bonding group was almost completely canceled by the counteracting effects - the push-back effect of the bonding group, and the dipole and push-back effects of the terminal group. Among these, the push-back effect of the terminal group can be omitted because the distance from the electrode surface is larger than the spatial extent to which the electrons spill away from the Au surface (~2.6 Å (Ref. 30)). From the switching direction of the BT device, the push-back effect of the thiol bonding group is larger than its dipole effect. In the NBT device, the $\Delta\Phi_m$ value obtained by subtracting that of the bonding group's dipole effect from that of the bonding group's push-back effect was roughly equal to the $\Delta\Phi_m$ of the terminal group's dipole effect. The very low $R_{sw}$ of the NBT device implies that the SAM molecules with oppositely oriented dipoles at bonding and terminal groups should be a good choice for achieving stable operation of organic electronic devices by using SAM-modified electrodes.

The $R_{sw}$ values of the MBT device were comparable to or somewhat lower than those of the BT device. The MBT's terminal group possesses a dipole pointing towards the methyl group from the benzene ring, which is in the same direction as the dipole of the thiol bonding group. Intuitively, the dipole at the terminal group should enhance the structural switching because the total force exerted by the external electric field increases. This effect should increase $R_{sw}$. However, the magnitude of methyl terminal group's dipole is 0.37 D, considerably smaller than that of the nitro group (4.01 D). Thus, the enhancement of the structural change might be limited. In addition, the addition of the methyl group is expected to increase the structural stability of the molecular monolayer through an enhanced van der Waals interaction between the molecules. This effect can be easily understood from the fact that the degree of ordering of alkanethiol-based SAMs increases for longer alkyl chains.[31] This should reduce $R_{sw}$. The rather comparable $R_{sw}$ values of the BT and MBT devices can be explained as resulting from the opposite effects of the enhancements in the structural stability and structural change.

**Modeling the switching strength.** The $\Delta\Phi_m$ that is induced by the dipole effect is expressed by Eq. (1). If the $\Delta\Phi_m$ that is induced by the push-back effect was known, the overall $\Delta\Phi_m$ could be evaluated. The $\Delta\Phi_m$ induced by the push-back effect is considered to be roughly proportional to the overlap of the two electronic clouds of the electrode and the molecule on it.[32] Thus, by using the average distance between the dipole at the bonding group, $z$, the $\Delta\Phi_m$ that is induced by the push-back effect can be expressed as:

$$\Delta\Phi_m^{\text{push-back}} = -CN\left(\frac{d}{2} - z\right) = -CN\frac{d}{2}[1-\cos(\theta)] = -CNd\sin^2\left(\frac{\theta}{2}\right), \tag{4}$$

where $C$ is the proportionality constant and $d$ is the distance between the dipole's positive and negative poles. The $\Delta\Phi_m^{\text{push-back}}$ always attains negative values. Strictly speaking, the $\Delta\Phi_m^{\text{push-back}}$ should include the bond dipole owing to the S-Au bonds, and it should be finite even when $\theta = 0$. However, the value of $R_{sw}$ depends only on how much the $\Phi_m$ changes (see Eq. (3)); when we consider $R_{sw}$, the bond dipole should be cancelled out because it can be regarded as independent of the change in $\theta$. Thus, the bond dipole is not included in the calculations below. In addition, the difference in the structural stability owing to the multiple bonding groups and the presence of the terminal group is not included in the calculations below. This can be justified because the calculation is performed as a function of $\theta$, and the structural stability only determines the extent of molecular tilting, *i.e.*, the possible range of the variation in $\theta$, $\delta\theta$. As shown in Eqs. (1) and (4), both $\Delta\Phi_m^{\text{dipole}}$ and $\Delta\Phi_m^{\text{push-back}}$ depend on $\theta$. Thus, it can be determined by how much the molecules are tilted after applying the $V_{sw}$.

The surface molecular densities were reported to be $3.2 \times 10^{14}$ cm$^{-2}$, $4.5 \times 10^{14}$ cm$^{-2}$ and $6.4 \times 10^{14}$ cm$^{-2}$ for the BT on Au(111),[33] MBT on Au(111),[34] and B2T on an evaporated Au film formed on glass,[35] respectively. Among these, the surface of the evaporated Au film was rough, and the actual surface area was larger than the film area. Thus, the actual density should be obtained by dividing by the roughness factor, which was reported to be 1.7 for evaporated Au on mica without annealing.[36] The actual density for the B2T can be calculated as $3.8 \times 10^{14}$ cm$^{-2}$ by using the roughness factor of 1.7. The surface density for the NBT is unknown; to a first approximation, $N$ will be considered to be the average value of the three surface molecular densities, $3.8 \times 10^{14}$ cm$^{-2}$, for all the monothiol BT derivatives treated in this study. The $N$ of the B2T device was set to twice the above value because the B2T possesses two thiol bonding groups.

The tilt angles of as-fabricated monolayers on Au, $\theta_0$, were reported to be 49°, 60°, and 51° for BT,[37] MBT,[38] and B2T,[35] respectively. Although angle $\theta_0$ is unknown for NBT, it will be taken as 53° for all BT derivatives, corresponding to the average of the three values given above. In the calculations below, it will be assumed that $\theta_0$ is the center value for the structural change of the DM molecules, i.e., $\theta_0 - \delta\theta/2 < \theta < \theta_0 + \delta\theta/2$. It should be noted that the surfaces of vacuum-deposited Au films used in the experiments in this study were not atomically smooth. However, the electric fields at the Au surfaces were normal to the surfaces. Thus, the tilt angle of the DM molecules and the direction of an external electric field can be treated as shown in Figure 6A.

The $\Delta\Phi_\mathrm{m}$ induced by the bonding group's dipole effect was calculated by using Eq. (1) with no adjustable parameter. The values of $\mu_0$ for the thiol bonding group were set to 0.86 D and 1.18 D for the monothiol and dithiol molecules, respectively. The value of $\varepsilon^\mathrm{eff}$ for the BT derivative monolayer was ca. 2.5.[25] Only the value of $Cd$ was unknown, and this value should have been selected for reproducing the experimentally obtained $R_\mathrm{sw}$ values. The value of $\delta\Phi_\mathrm{B}$ was extracted by using Eq. (3) from the $R_\mathrm{sw}$ data shown in Figure 3D; the $\delta\Phi_\mathrm{B}$ values for $|V_\mathrm{sw}|$ of 7 V were determined to be 90, 181, 63, and 19 mV for the BT, B2T, MBT, and NBT devices, respectively. The value of $Cd$ should be chosen for reproducing these $\delta\Phi_\mathrm{B}$ values. To obtain $Cd$ for the thiol bonding group, the value obtained by subtracting $\Delta\Phi_m$ at $\theta = \theta_0 + \delta\theta/2$ and at $\theta = \theta_0 - \delta\theta/2$ should be the same as the $\delta\Phi_B$ values.

To start the calculation, the $\delta\Phi_B$ value of the BT device (90 mV for $|V_{sw}| = 7$ V) was taken as a standard for comparison. The calculated $Cd$ values by using the $\delta\Phi_B$ value of 90 mV for various $\delta\theta$ are shown in Figure 6A. The determined $Cd$ values corresponding to the different $\delta\theta$ values for the BT were used for determining the $\delta\theta$ values for B2T by reproducing the experimentally obtained $\delta\Phi_B$ value of the B2T device (181 mV for $|V_{sw}| = 7$ V). In the calculation result shown in Figure 6B, the maximal value of the vertical axis was set to 74° because it must be less than twice the difference between the maximal allowed $\theta$ (90°) and $\theta_0 = 53°$. Therefore, the value of $\delta\theta$ for the BT should be below 23.6° that gives the maximal change in $\delta\theta$ for B2T.

Dithiol molecules are known to form DMs with higher disorder than their monothiol counterparts.[29] As a result, the B2T DM should have lower structural stability than the BT DM. Thus, the value of $\delta\theta$ for B2T should be larger than that for BT, which is clearly reproduced in Figure 6B. The center value of the possible range of $\delta\theta$ for BT (11.8°) was used for calculating the representative $\Delta\Phi_m$-$\theta$ characteristics for the BT and B2T devices (Figure 6C).

For the MBT and NBT, the effects of the terminal group have to be included in the calculation. To a first approximation, the push-back effect of the terminal group can be omitted because the distance from the electrode surface is considerably larger than that for the bonding group. When considering the dipole effect expressed by Eq. (1), the tilt angles of the thiol bonding group and terminal group should be different because $\mu_0$ differs between these two groups. The values of $\theta_0$ for the thiol bonding and terminal groups can be considered to be identical, but the possible ranges of $\theta$ variation should be different.

In the case of NBT, the dipoles at the bonding and terminal groups point in opposite directions. In addition, $|\mu_0|$ of the nitro terminal group (4.01 D) is much larger than that of the thiol bonding group (0.86 D). Thus, the structural change induced by the bonding group is considered to be encumbered by the opposite structural change induced by the terminal group. At present, the correlation between the structural changes induced by the bonding and terminal groups is unknown.

In the case of MBT, both dipoles at the bonding and terminal groups point towards the electrode surface. The $|\mu_0|$ of the methyl terminal group (0.37 D) is considerably smaller than that of the thiol bonding group (0.86 D). Thus, we presume that the correlation between the structural changes of these groups is negligible, and $\delta\theta$ of the terminal group is identical to that of the bonding group. The dipole-dipole interaction between the bonding and terminal group should vary upon tilting because the separation between these two groups changes, which should change the $\varepsilon^{eff}$ value. However, for a rather small dipole moment of the terminal group, the separation-dependent change in the bonding-terminal interaction was reported to be weak.[39] Thus, the possible $\theta$ dependence of $\varepsilon^{eff}$ is not considered here, which might not significantly affect the calculation results shown below. While keeping the condition prescribing that $Cd$ is determined for reproducing the $\delta\Phi_B$ value of the BT for $|V_{sw}|$ of 7 V (90 mV), the value of $\delta\theta$ was chosen to reproduce the $\delta\Phi_B$ value of the MBT for $|V_{sw}|$ of 7 V (63 mV). Figures 6B and 6C also show the calculated $\delta\theta$ values and $\Delta\Phi_m$-$\theta$ characteristics for MBT when $\theta_0 = 53°$ and $\delta\theta$ for BT was equal to 11.8°.

In the above calculations, the $Cd$ values for each $\delta\theta$ of the BT were first calculated for reproducing the $\delta\varPhi_B$ value of the BT device (90 mV for $|V_{sw}| = 7$ V), and then the obtained $Cd$ values were used for calculating the $\delta\theta$ values of the B2T and MBT that reproduced the $\delta\varPhi_B$ values of the B2T and MBT devices, respectively. To validate the calculation, the same procedure was started with calculating the $Cd$ values for each $\delta\theta$ of the B2T or MBT for reproducing the $\delta\varPhi_B$ values of the B2T (181 mV for $|V_{sw}| = 7$ V) and MBT devices (63 mV for $|V_{sw}| = 7$ V), respectively. The alternative calculation yielded quantitatively the same results as those in Figure 6A (Figure 7).

## **CONCLUSIONS**

Electrically stimulated switching of a charge injection barrier at electrode/organic semiconductor interfaces was investigated by using various BT derivatives as a DM on the electrode surfaces. The switching behavior was induced by the structural changes in the DM molecules, and was manifested as a reversible inversion of the polarity of the DM-modified Au electrode/rubrene /DM-modified Au electrode diodes. All of the tested BT derivatives exhibited the same switching direction regardless of the direction of the overall dipole of the derivatives. From this result, the push-back effect of the thiol bonding group was found to dominantly determine the switching direction, while the terminal group modulated the switching strength. A device with B2T DMs exhibited the highest switching ratio, confirming the effectiveness of the strategy employing DMs with multiple bonding groups. The switching ratios of the B2T device were ca. 20, $10^2$ and $10^3$ for $|V_{sw}|$ of 3, 5 and 7 V, respectively. These ratios corresponded to the changes in the hole-injection barrier heights of 75, 114 and 181 meV. These results demonstrate that the molecular switching device investigated here can be operated at low voltages.

Model calculations of the change in the charge injection barrier height were also performed. A variation in $\theta$ of the BT device owing to the $V_{sw}$ application was estimated to be smaller than 23.6°. To verify the calculation results, further studies are needed for determining the $\theta$ values by using direct measurement methods such as the surface X-ray scattering technique.[40]

The present study has unveiled the switching nature of the electrode modification layers. Such surface modification has been widely employed in organic electronic devices such as field-effect transistors and light-emitting diodes. Furthermore, unintentional contaminations can easily adsorb on the electrodes of organic-on-electrode type interfaces. The electrically stimulated switching of the charge injection barrier height directly leads to the instability of the operation of these devices.[41] The understanding offered by the present study can be exploited for obtaining highly stable operations of organic electronic devices, especially with molecular modification layers.

**METHODS**

A highly doped Si wafer with a thermally grown 300-nm-thick oxide layer on top of it was used as a substrate for the device fabrication. The substrate was cleaned with acetone and isopropanol by using an ultrasonic bath, and was then dried by using an air blower. Au electrodes with inter-electrode spacing of 0.4 μm were fabricated by conventional electron-beam lithography. The Au electrodes were 14-nm-thick, and a thin 1-nm-thick Cr layer was formed underneath the Au layer for strengthening the adhesion of the Au film to the underlying $SiO_2$ substrate.

After the electrode formation, the substrate was exposed to oxygen plasma to remove contaminants such as residues of an electron-beam resist. Immediately after the oxygen plasma cleaning, the substrate was immersed in pure ethanol for 30 min to reduce slightly oxidized Au electrodes.[42]

Then, the substrate was immersed in a 1-mM solution of the BT derivative, and the air in the solution container was replaced by Ar immediately. The immersion time and temperature were 24 h and room temperature, respectively. BT derivatives are known to form a DM without long-range ordering on Au surfaces at room temperature.[43-45] After the 24-h-long immersion, ultrasonication in the pure solvent was performed to remove physisorbed layers of the BT derivative, which left a chemisorbed monolayer on the Au electrodes. Ethanol and tetrahydrofuran were used as the solvent for the monothiol molecules (BT, MBT and NBT) and the dithiol molecule (B2T), respectively.

Finally, a rubrene single crystal grown by the physical vapor transport[46] was laminated onto the electrodes. The growth process was repeated three times to increase the crystal's purity; the freshly synthesized crystal was manually scooped by a hair attached to a stick, and transferred onto the substrate with the DM-modified Au electrodes under optical microscope.

Electrical measurements were performed in ambient air, under ambient light, at room temperature.

## ASSOCIATED CONTENT

**Supporting Information**. Average $R_{sw}$ as a function of $|V_{sw}|$ for the device with no modification layer. This material is available free of charge via the Internet at http://pubs.acs.org.

## AUTHOR INFORMATION

**Corresponding Author**

*Address correspondence to r-nouchi@21c.osakafu-u.ac.jp.

**Notes**

The authors declare no competing financial interest.

## ACKNOWLEDGMENT

This work was supported in part by the Special Coordination Funds for Promoting Science and Technology from the Ministry of Education, Culture, Sports, Science and Technology of Japan; a Grant-in-Aid for Challenging Exploratory Research (No. 25600078) from the Japan Society for the Promotion of Science; and the Murata Science Foundation.

## ABBREVIATIONS

BT, benzenethiol; DM, disordered monolayer; HOMO, highest occupied molecular orbital; MBT, 4-methylbenzenethiol; NBT, 4-nitrobenzenethiol; B2T, 1,2-benzenedithiol; SAM, self-assembled molecular monolayer.

Figures:

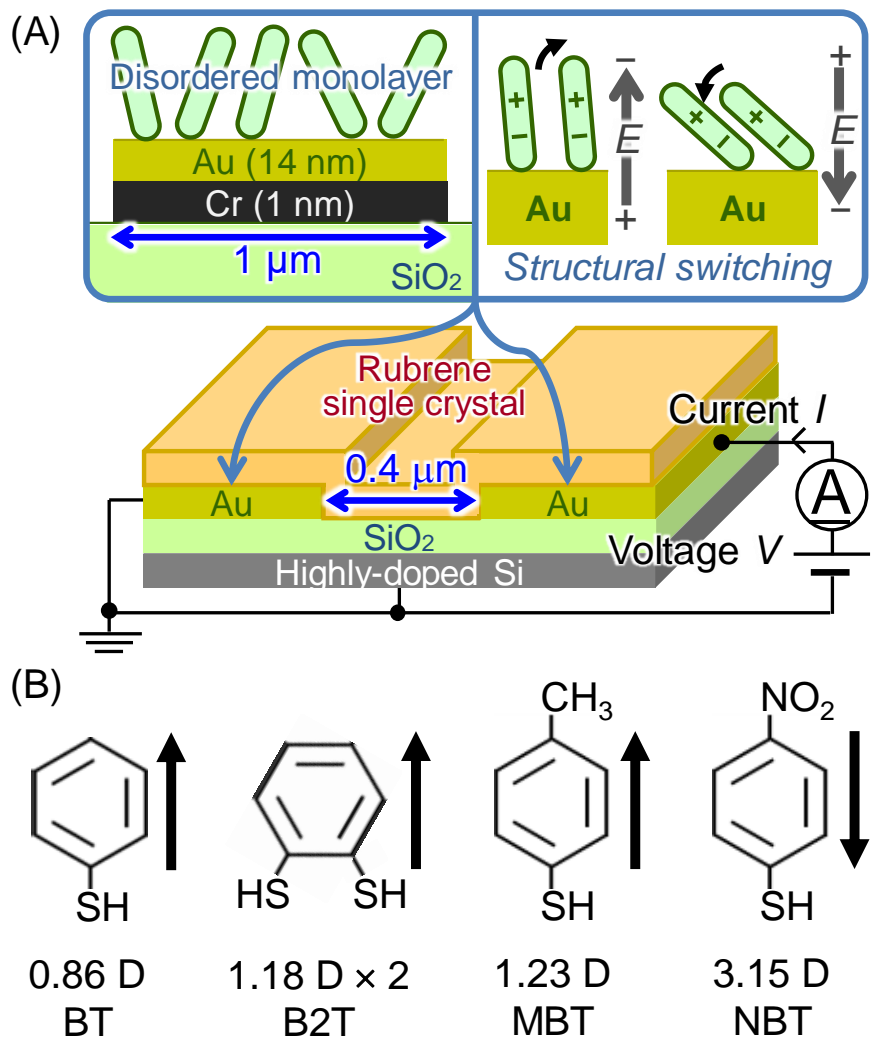

**Figure 1.** (A) Schematic of the switching device tested in this study. $E$ denotes the external electric field. (B) Molecular structures of the molecules comprising the electrode modification layer. The magnitudes of the permanent electric dipoles along the surface normal direction are shown.

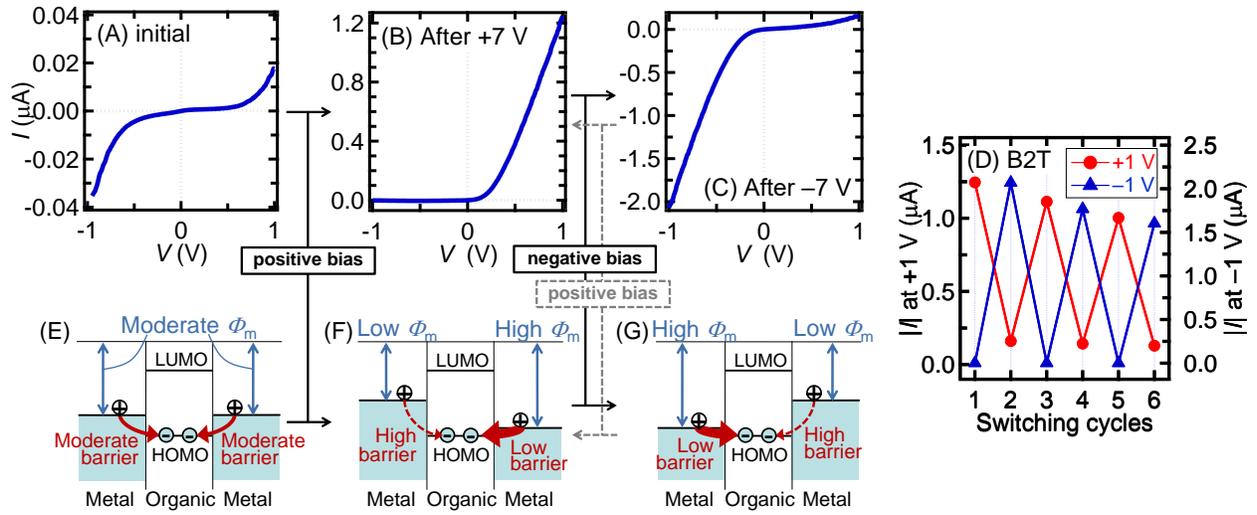

**Figure 2.** *I-V* characteristics of a device with electrodes modified with B2T monolayers, measured immediately after (A) device fabrication, (B) application of $V_{sw} = +7$ V for ~60 s, and (C) application of $V_{sw} = -7$ V for ~60 s. (D) Switching cycle of the absolute current $|I|$ with respect to the consecutive application of $V_{sw} = \pm 7$ V. (E–G) Schematic band diagrams deduced from (A–C), respectively.

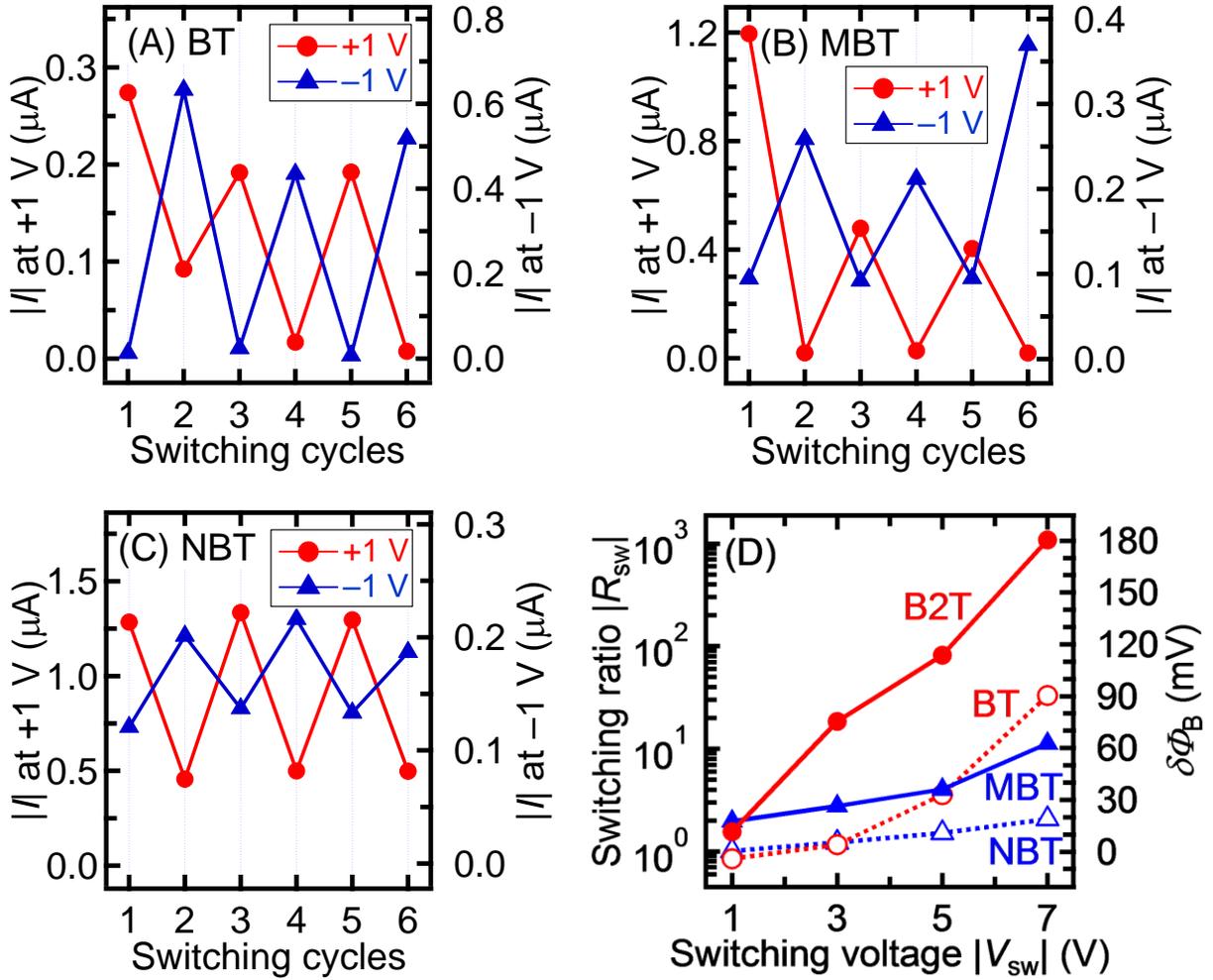

**Figure 3.** Switching cycle of the absolute current |*I*| with respect to the consecutive application of $V_{sw} = \pm 7$ V for devices with electrodes modified with (A) BT, (B) MBT, and (C) NBT monolayers. (D) Average $R_{sw}$ as a function of $|V_{sw}|$. The change in the barrier height, $\delta\Phi_B$, is also calculated by using Eq. (3) from the $R_{sw}$ data.

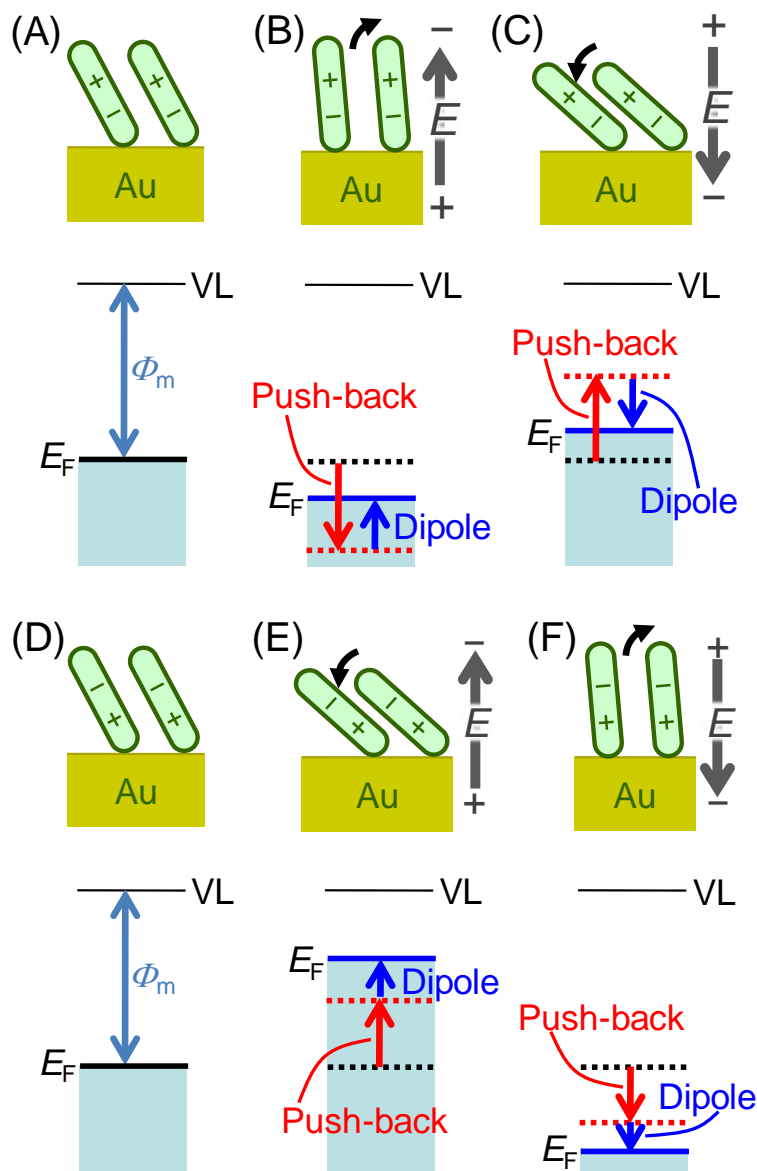

**Figure 4.** Structure of the molecular monolayer and the expected energy diagrams. The $\varDelta\varPhi_\mathrm{m}$ owing to the dipole and push-back effects are considered in the schematics. VL denotes the vacuum level. (A–C) DM molecules with permanent electric dipoles pointing outward from the electrode surface (BT, MBT, and B2T): (A) as fabricated, (B) after applying an upward electric field, and (C) after applying a downward electric field. (D–F) Same as in (A–C), but DM molecules with dipoles in the opposite direction (NBT).

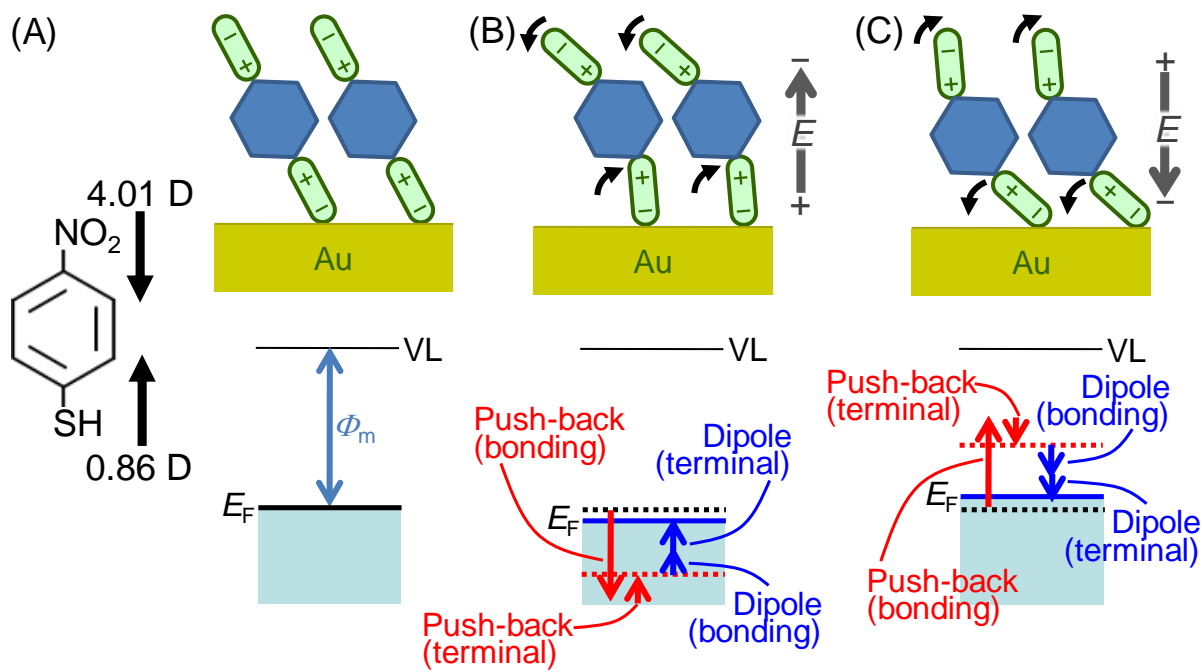

**Figure 5.** Detailed structure of the NBT monolayer and its corresponding energy diagrams (A) as fabricated, (B) after applying an upward electric field, and (C) after applying a downward electric field. The C-N bond at the terminal group is assumed to possess a certain degree of flexibility in (B) and (C).

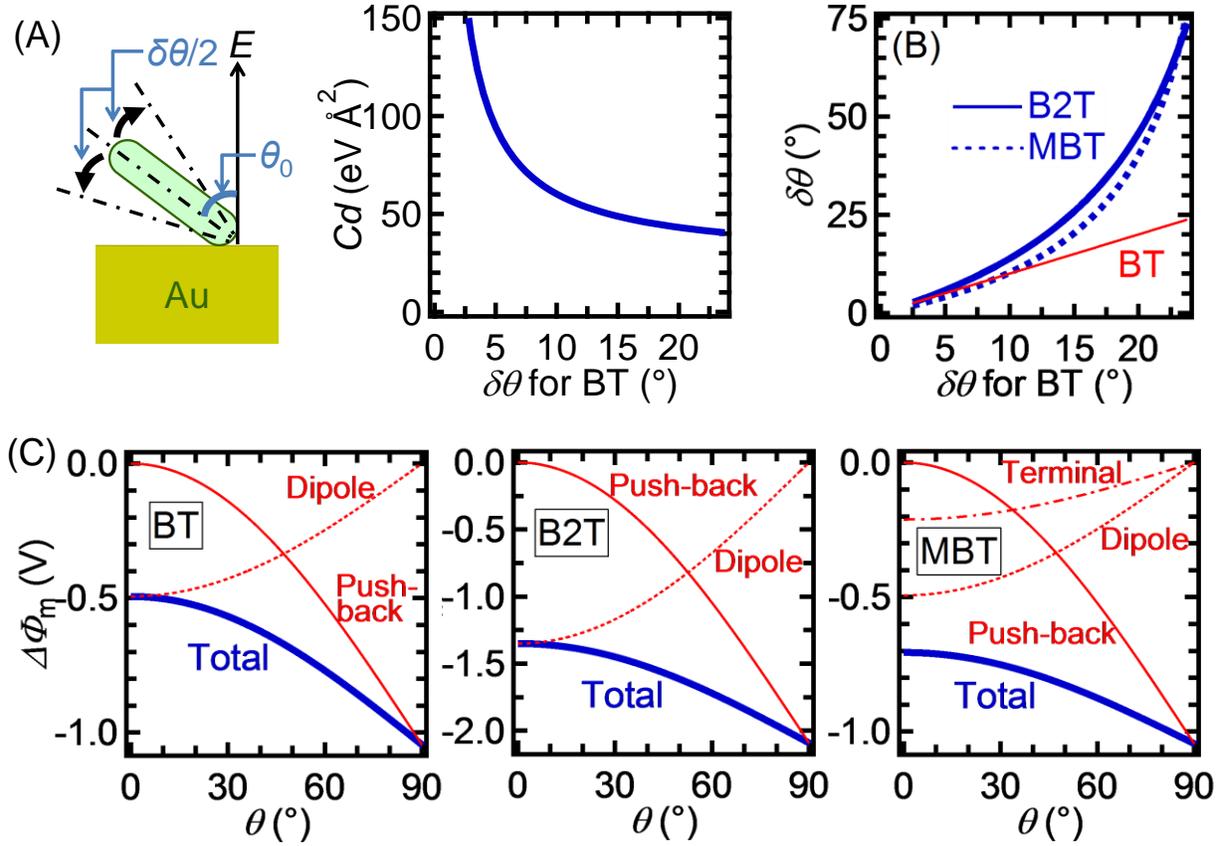

**Figure 6.** Calculation of $\Delta\Phi_m$ for BT, B2T and MBT monolayers. (A) Schematic of the tilt angle (left) and calculated $Cd$ values that reproduce $\delta\Phi_B$ value of 90 mV for the BT monolayer (right). $Cd$ values were determined for the condition in which the value obtained by subtracting $\Delta\Phi_m$ at $\theta = \theta_0 + \delta\theta/2$ from $\Delta\Phi_m$ at $\theta = \theta_0 - \delta\theta/2$ became the same as the $\delta\Phi_B$ value. Angle $\theta_0$ was set to 53°. (B) Calculated $\delta\theta$ values for B2T and MBT, reproducing the $\delta\Phi_B$ values of 181 and 63 mV for B2T and MBT, respectively. $Cd$ values shown in (A) were used. (C) Calculated $\Delta\Phi_m$ for the BT, B2T, and MBT, for which $\delta\theta$ of BT was set to 11.8°.

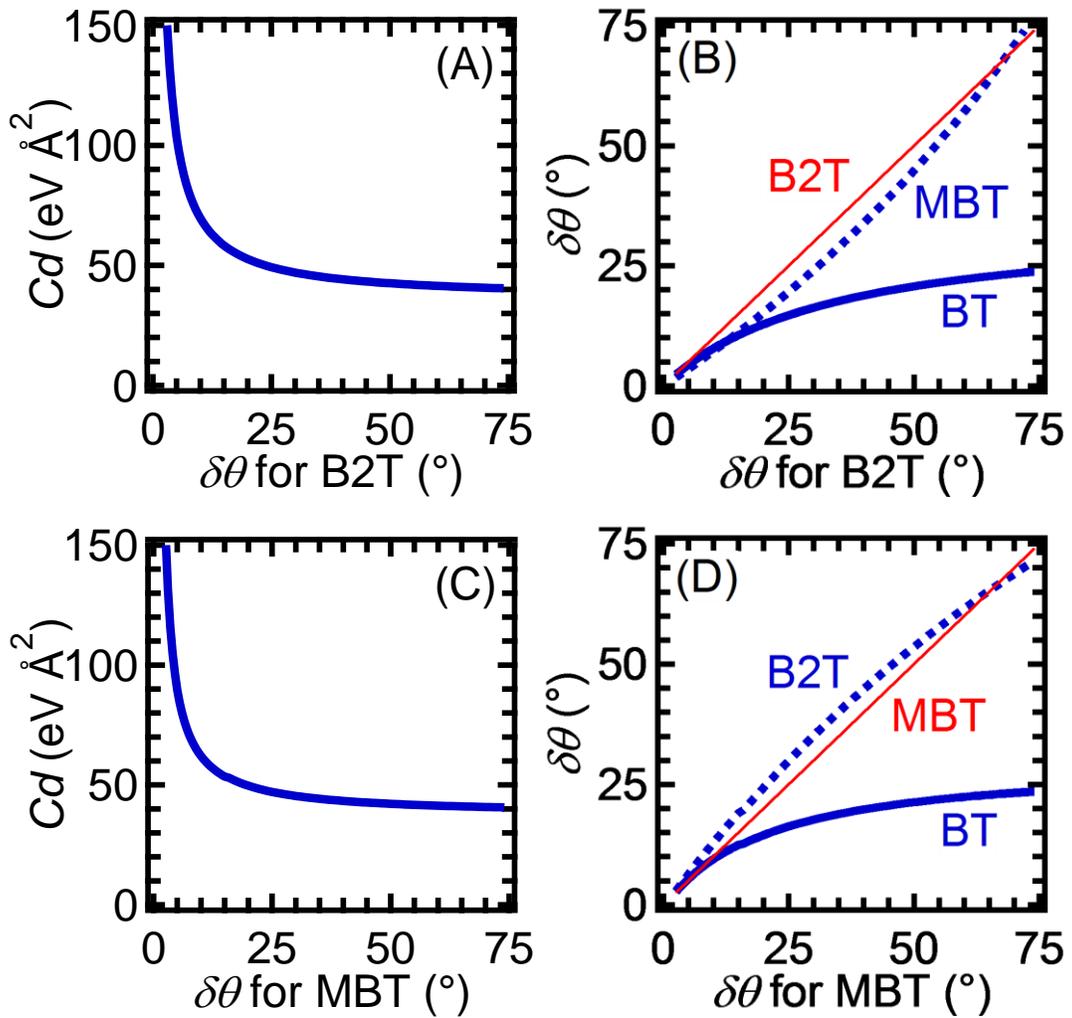

**Figure 7.** Values of $Cd$ and $\delta\theta$ calculated by alternative procedures. (A) Calculated $Cd$ values that reproduce $\delta\Phi_B$ value of 181 mV for the B2T. $Cd$ values were determined under the condition in which the value obtained by subtracting $\Delta\Phi_m$ at $\theta = \theta_0 + \delta\theta/2$ from $\Delta\Phi_m$ at $\theta = \theta_0 - \delta\theta/2$ became the same as the $\delta\Phi_B$ value. Angle $\theta_0$ was set to 53°. (B) Calculated $\delta\theta$ values for BT and MBT that reproduce $\delta\Phi_B$ values of 90 and 63 mV for BT and MBT, respectively. $Cd$ values shown in (A) were used. (C, D) Calculated $Cd$ and $\delta\theta$ values obtained by using the same procedure in which $Cd$ was first calculated for reproducing $\delta\Phi_B$ value of 63 mV for MBT.

Supporting Information for

# Substituent-Controlled Reversible Switching of Charge-Injection-Barrier Heights at Metal/Organic-Semiconductor Contacts Modified with Disordered Molecular Monolayers

*Ryo Nouchi\* and Takaaki Tanimoto*

Nanoscience and Nanotechnology Research Center, Osaka Prefecture University, Sakai 599-8570, Japan

Contents:

- The device with no modification layer on the electrodes

**The device with no modification layer on the electrodes**

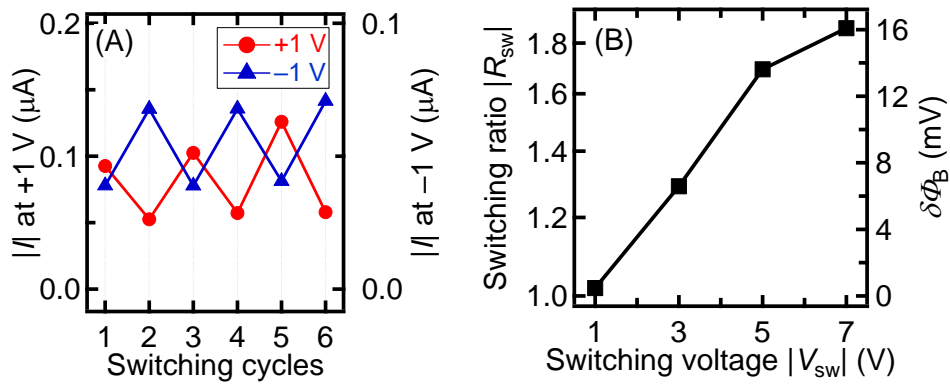

**Figure S1.** (A) Switching cycle of the absolute current $|I|$ with respect to the consecutive application of $V_{sw} = \pm 7$ V, and (B) the average $R_{sw}$ as a function of $|V_{sw}|$ for the device with no modification layer. The change in the barrier height, $\delta\Phi_B$, is also calculated by using Eq. (3) from the $R_{sw}$ data.

Figure S1 shows the switching cycle and the $R_{sw}$-$V_{sw}$ characteristics of the device with no modification layer intentionally formed onto the Au electrodes. The $R_{sw}$ values are lower than those of the B2T, BT and MBT devices (see Figure 3 in the main text), confirming that the higher $R_{sw}$ values of these devices with modification layers originate from the presence of the modification layers. By contrast, the $R_{sw}$ values are comparable to those of the NBT device. However, as explained in the next paragraph, this similarity is considered as coincidental.

For formation of the modification layers, the following steps were carried out: (1) cleaning of as-fabricated Au electrodes by oxygen plasma, (2) reduction of the slightly oxidized Au surface by immersion in ethanol, and (3) immersion in a 1 mM solution of each BT derivatives. The second and third steps were performed immediately following the corresponding preceding step. Thus, the molecular layer on the electrode surface should be exactly the BT derivative monolayer. Thus, the switching characteristics of the NBT device are considered to be the characteristics of the NBT layer. On the other hand, the third step was not performed for fabrication of devices with no modification layer. In this case, the surface of the reduced Au electrodes was contaminated by hydrocarbon molecules in the air during the lamination of rubrene crystals in ambient air. Thus, the observed weak switching of the device without intentionally formed monolayers is considered to originate from the unintentionally formed contamination layer.